\documentclass[draftcls,onecolumn,12pt]{IEEEtran}

\usepackage[dvips]{graphicx}
\usepackage[cmex10]{amsmath}
\usepackage{subfigure,cite,bbm,amssymb,multirow,amsthm,bm}
\usepackage{dsfont}
\usepackage{url}
\usepackage{stfloats}
\usepackage{color}
\usepackage[noend]{algpseudocode}
\usepackage{algorithm}
\usepackage{flushend}
\usepackage{pgfplots}
\usepackage{tikz}

\makeatletter
\newlength \figwidth
\if@twocolumn
\else
\fi
\makeatother

\providecommand{\abs}[1]{\lvert#1\rvert}												
\renewcommand{\b}[1]{\ensuremath{\mathbf{#1}}}		 							
\newcommand{\bs}[1]{\ensuremath{\boldsymbol{#1}}}		 						
\renewcommand{\c}[1]{\ensuremath{\mathcal{#1}}} 								
\newcommand{\E}[1]{\ensuremath{\mathbb{E}\left[#1\right]}} 		  
\newcommand{\Es}[1]{\ensuremath{\mathbb{E}^{*}\left[#1\right]}} 
\newcommand{\mat}[1]{\ensuremath{\begin{bmatrix}#1\end{bmatrix}}}
\newcommand{\mr}[1]{\ensuremath{\mathring{#1}}}	

\def \y {{\b{y}}}
\def \yt {{\b{y}(t)}}
\def \ypt {{y_p(t)}}
\def \yst {{\b{y}_s(t)}}
\def \by {{\bar{\b{y}}_s(t)}}

\def \ybt {{\b{y}_{\bar{s}}(t)}}
\def \yh {{\hat{\b{y}}_{\bar{s}}}}
\def \chit {{\bs{\chi}(t)}}
\def \chih {{\hat{\bs{\chi}}}}
\def \bnu {{\bs{\nu}}}
\def \nut {{\bs{\nu}(t)}}

\def \nust {{\bs{\nu}_s(t)}}

\def \nubt {{\bs{\nu}_{\bar{s}}(t)}}

\def \epst {{\bs{\epsilon}_s(t)}}
\def \epsbt {{\bs{\epsilon}_{\bar{s}}(t)}}
\def \et {{\bs{\eta}(t)}}
\def \xt {{\b{x}(t)}}
\def \Exp {\mathbb{E}}

\def \tc {{\tilde{\bs{\chi}}(t)}}
\def \My {{\b{M}^{\b{y}}_{\bar{s}}(t)}}
\def \Kt {{\b{K}(t)}}
\def \bbeta {{\bs{\beta}}}

\def \ceta {{\b{C}_{\bs{\eta}}}}

\def \si  {{\sigma^2\b{I}_S}}
\def \sib  {{\sigma^2\b{I}_{P-S}}}
\def \cnu {{\b{C}_{\bs{\nu}}}}

\def \P {{\c{P}}}
\def \S {{\c{S}}}
\def \N {{\c{N}}}
\def \A {{\c{A}}}
\def \B {{\c{B}}}
\def \cSt {{\c{S}(t)}}
\def \cSb {{\bar{\c{S}}(t)}}
\def \St {{\b{S}(t)}}
\def \Stt {{\b{S}^T(t)}}
\def \Sb {{\bar{\b{S}}(t)}}
\def \Sbt {{\bar{\b{S}}^T(t)}}
\def \sbr {{\bar{s}}}

\theoremstyle{plain}

\newtheorem{prop}{Proposition}

\theoremstyle{definition}

\theoremstyle{remark}
\newtheorem{rem}{\bf Remark}

\begin{document}
\title{Dynamic Network Delay Cartography}

\author{\emph{Ketan Rajawat, Emiliano Dall'Anese, and Georgios B. Giannakis}$^\star$

\thanks{Submitted April 19, 2012; revised \today.}
\thanks{\protect\rule{0pt}{0.5cm}%
The authors are with the Department of Electrical and Computer Engineering, University of Minnesota, 200 Union Street SE, %
Minneapolis, MN 55455, USA. Tel/fax: +1(612)624-9510/625-2002. %
E-mails: {\tt \{ketan,emiliano, georgios\}@umn.edu}
$^\star$Corresponding author.
}
\thanks{\protect\rule{0pt}{0.5cm}%
Work in this paper was supported by NSF-ECCS grant no. 1202135. Part of this paper has been presented at the \emph{IEEE Statistical Signal Processing Workshop}, Ann Arbor, MI, Aug. 2012.}
}

\markboth{IEEE TRANSACTIONS ON INFORMATION THEORY (SUBMITTED)}{}

\maketitle

\begin{abstract}
Path delays in IP networks are important metrics, required by network operators for assessment, planning, and fault diagnosis.
Monitoring delays of all source-destination pairs in a large network is however challenging and wasteful of resources.
The present paper advocates a spatio-temporal Kalman filtering approach to construct network-wide delay maps using measurements on only a few paths. 
The proposed network cartography framework allows efficient tracking and prediction of delays by relying on both topological as well as historical data.
Optimal paths for delay measurement are selected in an online fashion by leveraging the notion of submodularity.
The resulting predictor is optimal in the class of linear predictors, and outperforms competing alternatives on real-world datasets.
\end{abstract}


\newpage

\section{Introduction}\label{intro}

The explosive growth in network traffic volumes has necessitated the development of avant-garde monitoring tools to endow network operators with a comprehensive view of the global network behavior.  However, acquisition and processing of network-wide performance metrics for large networks is no easy task.  For instance, monitoring path metrics such as  delays or loss rates is challenging primarily because the number of paths generally grows as the square of the number of nodes in the network. Therefore, measuring and storing the delays of all possible source-destination pairs is hard in practice even for moderate-size networks.

Focus has thus shifted towards statistical means of predicting network-wide performance metrics using measurements on only a subset of nodes \cite{mich11,nowak11}.
A promising approach in this context has been the application of \emph{kriging}, a tool for spatial prediction popular in geostatistics and environmental sciences \cite{Rip81,Cressie}.  
A \emph{network kriging} approach was developed in \cite{nk}, where  network-wide path delays were predicted using measurements on a chosen subset of paths. 
The class of linear predictors introduced leverages network topology information to model the covariance among path delays.
This is accomplished in \cite{nk} by assigning higher correlation between two paths if they share several links, as in this case, they are expected to incur similar delay variations.

The present paper puts forth a \emph{dynamic} network kriging approach capable of real-time spatio-temporal delay predictions.
Specifically, a kriged Kalman filter (KKF) is employed to explicitly capture variations due to queuing delays, while retaining the topology-based kriging predictor.
The resulting dynamic network kriging approach not only yields lower prediction error, but is also more flexible, allowing delay measurements to be taken on random subsets of paths.
In this context, the problem of choosing the optimal paths for delay measurements is also considered. 
Since the KKF runs in real-time, the paths are also selected in an online fashion by minimizing the prediction error per time slot. 
Interestingly, the resulting combinatorial optimization problem is shown to be submodular, and is therefore solved approximately via a greedy routine. 

Recently, a compressive sampling-based approach has also been reported for predicting network-wide performance metrics \cite{coates,xu}. 
For instance, diffusion wavelets were utilized in \cite{coates} to obtain a compressible representation of the delays, and account for spatial and temporal correlations.
Although this allows for enchanced prediction accuracy over \cite{nk}, it requires batch processing of measurements which does not scale well to large networks for real-time operation. 
In contrast, both the KKF and the greedy path selection algorithms entail sequential operations, and are therefore significantly faster.

Imputation of end-to-end delays has also been considered in the context of Internet geolocation. 
Treating end-to-end delays as distances between nodes, all-pair node distances are estimated using Euclidean embedding \cite{vivaldi}, or, matrix factorization \cite{fact11}. 
However, these approaches do not exploit the temporal or topological information, since their focus is not on monitoring or extrapolation (that is, prediction) of delays.

The rest of the paper is organized as follows. Sec. \ref{probstat} introduces the model and the problem statement. Sec. \ref{kriging} deals with the KKF approach, while Sec. \ref{estparam} describes techniques for estimating the relevant parameters. 
Finally, empirical validation of KKF and comparisons with the Kriging approach of \cite{nk} 
are provided in Sec. \ref{ev}.  


\emph{Notation}. Lower case symbols with indices, such as $y_p$, represent scalar variables. These variables, when stacked over their indices are denoted through their bold-faced versions $\b{y}$. Bold-faced upper case symbols ($\b{S}$) represent matrices. Regular upper case symbols ($S$) represent constant scalars, and typically stand for the cardinality of the set represented by corresponding calligraphic upper case symbol ($\c{S}$). Identity matrix of size $P \times P$ is denoted by $\b{I}_P$, , and its columns by $\b{e}_1$, $\b{e}_2$, $\ldots$, $\b{e}_P$. Matrix $\b{C}_{\b{y}}$ denotes the covariance matrix of the vector $\b{y}$. 

\section{Modeling and Problem Statement} \label{probstat}

Consider an IP network modeled by a connected digraph $\c{G} = (\c{V},\c{E})$, with $\c{V}$ denoting the set of nodes (devices, servers, or routers), and $\c{E}$, the communication links.
The issue is to monitor path delays on a set of multi-hop paths $\P$ that connect the $P:=\abs{\P}$ source-destination pairs. 
Latency measured on path $p \in \P$ at time $t$ is denoted by $\ypt$, and all such network-wide delays are collected in the vector $\b{y}(t)$.
At any time $t$ however, delay can only be measured on a subset of paths $\c{S}(t)\subset \P$, which is represented by $\b{y}_{s}(t)$.
Based on such partial current and past measurements $\c{H}(t) := \{\b{y}_{s}(\tau)\}_{\tau=1}^t$, the goal is to predict the remaining path delays $\b{y}_{\bar{s}}(t):= \{\ypt\}_{p\in \P\setminus\c{S}(t)}$ for each $t$.

The per-path end-to-end delay $\ypt$ consists of several independent components corresponding to contributions from each intermediate link and router. 
Of these, the queuing delay $\mr{\chi}_p(t)$ is the time spent by the packets waiting in the queues of intermediate buffers, and depends on the traffic volumes in competing links. 
{ Network traffic is not only correlated spatio-temporally, but also exhibits non-stationarities, in the form of random fluctuations and bursts \cite{structural}.
Indeed, it is not surprising that the statistical properties of queueing delays in large IP networks are largely unknown.
In the interests of model parsimony and amenability to the tools used later, the following random-walk model is instead adopted for the latent vector of queuing delays,
\begin{align}\label{eq:xi}
\mr{\bs{\chi}}(t) = \mr{\bs{\chi}}(t-1) + \bs{\eta}(t)
\end{align}
where $\bs{\eta}(t)$ denotes state noise with zero mean, and covariance matrix $\ceta:=\E{\bs{\eta}(t) \bs{\eta}^T(t)}$. Observe that the random-walk model has very few tuning parameters, compared to say, a model which includes a non-identity state transition matrix (i.e., $\mr{\bs{\chi}}(t) = \bs{B}\mr{\bs{\chi}}(t-1) + \bs{\eta}(t)$). Further advantages of the random-walk model, including those pertaining to the computational cost, are provided in later sections.}

Other components of the path delay, combined in the nonzero-mean random $\mr{\nu}_p(t)$, include the propagation, processing, and transmission delays, which are temporally uncorrelated (see e.g., \cite{bovy} for details).
This component of delays is however spatially correlated across paths, and the covariance matrix of the compacted vector $\mr{\bs{\nu}}(t)$ is given by $\cnu$.
Finally, the measurement of path delays using software tools such as \texttt{ping} itself introduces errors $\epsilon_p(t)$, which are assumed zero mean, uncorrelated over time and across paths, with covariance $\sigma^2:=\E{\epsilon_p(t)\epsilon_p^T(t)}$.

The measured delays are expressed as
\begin{align}
\ypt &= \mr{\chi}_p(t) + \mr{\nu}_p(t) + \epsilon_p(t) & p \in \c{S}(t). \nonumber
\end{align}
Letting $\St$ denote the $\abs{\c{S}(t)} \times P$ selection matrix with 0-1 entries that contains the $p$-th row of $\b{I}_{P}$ if $p \in \c{S}(t)$, the measurement equation can be compactly written as
\begin{align}
\label{eq:meas0}
\y_{s}(t) &= \St\mr{\bs{\chi}}(t) + \mr{\bs{\nu}}_s(t) + \epst
\end{align}
where the vector $\epst$ collects the measurement errors on paths $p \in \c{S}(t)$, and $\mr{\bs{\nu}}_s(t):=\St\mr{\bs{\nu}}(t)$. 

The next section describes a KKF approach for tracking and predicting the unknown end-to-end delays $\y_{\bar{s}}(t)$, by utilizing the state-space model described by \eqref{eq:xi} and \eqref{eq:meas0}. 
\section{Dynamic network kriging} \label{kriging}
The spatio-temporal model in \eqref{eq:xi}-\eqref{eq:meas0} is widely employed in geostatistics and environmental science, where $\mr{\bs{\chi}}(t)$ is generally referred to as trend, and $\mr{\bs{\nu}}(t)$ captures random fluctuations around $\mr{\bs{\chi}}(t)$; see e.g.,~\cite[Ch.~4]{Rip81},~\cite{MGRA98,WiC99}. 
Recently, a similar modeling approach was employed by~\cite{kim} to describe the dynamics of wireless propagation channels, and in~\cite{Cor09} for spatio-temporal random field estimation. 
{ In order to better relate the proposed model with the existing ones, the mean of $\mr{\bs{\nu}}(t)$ is incorporated in the trend, and \eqref{eq:meas0} is now replaced with
\begin{align}
\label{eq:meas}
\y_{s}(t) &= \St\bs{\chi}(t) + \bs{\nu}_s(t) + \epst
\end{align}
where $\bs{\nu}_s(t) := \mr{\bs{\nu}}_s(t) - \E{\mr{\bs{\nu}}_s(t)}$ and $\bs{\chi}(t) := \mr{\bs{\chi}}(t) + \E{\mr{\bs{\nu}}(t)}$, and likewise for $\bs{\nu}(t)$.}
Next, given only first- and second-order moments of $\et$, $\epst$, and $\bs{\nu}(t)$, this section derives the best linear predictor for the unavailable path delay vector $\y_{\bar{s}}(t)$.

Suppose first that the queuing delay vector $\bs{\chi}(t)$ is known, and let $\Sb$ denote an $\abs{\cSb}\times P$ matrix containing the $p$-th row of $\b{I}_P$ if $p \in \Sb$; that is, $\Sb$ is a path selection matrix which returns quantities pertaining to paths in $\Sb$.
Then, the linear minimum mean-square error (LMMSE) estimator (denoted by $\Es{.}$) for $\bnu_\sbr(t)$ is given by (see, e.g. \cite{AnM79})
\begin{align} \label{eq:kriging}
\Es{\bnu_\sbr(t) | \chit, { \yst}} &= \Sb\cnu\Stt\left(\St\cnu\Stt + \sigma^2\b{I}_S\right)^{-1}\nonumber\\
&\hspace{1cm}\times \left[\yst - \St\chit\right]
\end{align}
and is commonly referred to as kriging~\cite{Cressie}. 
In practice however, the trend $\bs{\chi}(t)$ has to be estimated from the data. 
In the so-termed universal kriging predictor~\cite{Rip81}, $\chit$ is estimated using the generalized least-squares (GLS) criterion, where $\nust$ is treated as noise (lumped together with $\epst$). The prediction for $\nubt$ is then obtained by replacing $\bs{\chi}(t)$ in \eqref{eq:kriging} with its estimate. This approach was proposed for network delay prediction in \cite{nk}, and was referred to as network kriging. 
However, since the trend is estimated independently using GLS per time slot, its temporal dynamics present in \eqref{eq:xi} are not exploited. 

From the spatio-temporal model set forth in Sec.~\ref{probstat}, it is clear that estimating the trend $\bs{\chi}(t)$ can benefit from processing both present and past measurements jointly. 
Towards this end, the Kalman filtering (KF) machinery offers a viable option for tracking the evolution of $\bs{\chi}(t)$ from the set of historical data $\c{H}(t)$. 
At each time $t$, the KF finds the LMMSE estimate $\chih(t) := \Es{\chit | \c{H}(t)}$, and its error covariance matrix $\b{M}(t) := \E{(\chit-\chih(t))(\chit-\chih(t))^T}$ using the following set of recursions (see e.g.,~\cite[Ch.~3]{AnM79})
\begin{subequations}\label{kkfeq}
\begin{align}
\chih(t) &= \chih(t-1) + \Kt(\yst - \St\chih(t-1))\label{iterx}\\
\b{M}(t) &= (\b{I}_P - \Kt\St)(\b{M}(t-1)+\ceta)\label{iterM}
\end{align}
\end{subequations}
where the so-termed Kalman gain $\Kt$ is given by
\begin{align}\label{gain}
\Kt &:= (\b{M}(t-1)+\ceta) \Stt \nonumber\\
&\hspace{-0.5cm}\times \left[\St(\cnu +\ceta + \b{M}(t-1))\Stt + \si \right]^{-1}.
\end{align}
Once $\chih(t)$ has been estimated via KF, $\nubt$ can be readily obtained via kriging as in \eqref{eq:kriging}, yielding the predictor
 {
\begin{align}\label{eq:kkf_estimate}
\yh(t) &= \Sb\chih(t) + \Sb\cnu\Stt\left(\St\cnu\Stt + \sigma^2\b{I}_S\right)^{-1}\nonumber\\
&~~~~\times \left[\yst - \St\chih(t)\right].
\end{align}
}
The predictor in \eqref{eq:kkf_estimate} constitutes what is also referred to as the kriged Kalman filter~\cite{MGRA98,WiC99}. 
The LMMSE framework employed here yields the best linear predictor even for non-Gaussian distributed noise. The prediction error of the KKF is characterized in the following proposition, whose proof is provided in Appendix \ref{Aerr}.
\begin{prop}
The prediction error covariance matrix at time $t$ is given by
\begin{subequations}
\begin{align}\label{errormat}
\My  &:= \Exp\{(\ybt-\yh(t))(\ybt-\yh(t))^T\} \\
&=\sigma^2\b{I}_{\bar{S}} + \Sb\left[(\b{M}(t-1) + \cnu + \ceta)^{-1} + \frac{1}{\sigma^2}\Stt\St\right]^{-1}\Sbt\,.
\end{align}
\end{subequations}
\end{prop}
Having a closed-form expression for the prediction error will come handy for selecting the matrix $\b{S}(t)$, as shown later in Sec. \ref{expdesign}.

The KF step also allows $\tau$-step prediction for $\tau\geq 1$, which is given by $\hat{\y}(t+\tau) = \chih(t)$, since the kriging term is temporally white. 
In the present context, this can be useful in preemptive routing and congestion control algorithms, as well as for extrapolating missing measurements. 
In the latter case, the covariance matrix is updated simply as $\b{M}(t) = \b{M}(t-1)+\ceta$.
Before concluding the description of the KKF, the following remarks are due.
\begin{rem}
The random walk model adopted in \eqref{eq:xi} may result in an unstable filter. Operationally, if the KKF is unstable, an incorrect initialization of $\b{M}(0)$ or $\bs{\chi}(0)$ may result in poor prediction performance even as $t \rightarrow \infty$.
This can be remedied by adopting a damped model $\bs{\chi}(t) = b\bs{\chi}(t-1) + \bs{\eta}(t)$ with $b < 1$. { Here, $\bs{\chi(t)}$ is a zero-mean random process which does not incorporate the mean of $\mr{\bs{\nu}}(t)$. The mean delay of all paths should instead be estimated a priori, and subtracted from the measurements themselves, so that each component of the path delay in \eqref{eq:meas} is zero-mean.}
With this modification, the results in this paper can be generalized to the damped case. 
The random walk model is nevertheless used here since no instability issues were observed in the two data sets considered in Sec. \ref{ev}.
{ An alternative formulation, developed along the lines of \cite{casas}, can also be used in the AR(1) case. This technique may however increase the number of state-space parameters, and considerably complicate the expressions developed in Sec. \ref{expdesign}.}
\end{rem}

\begin{rem}
A distributed implementation of the KKF may be desirable for enhancing the robustness and scalability of delay monitoring. 
In large-scale networks, a distributed algorithm also mitigates the message passing overhead required to collect all measurements at a fusion center. 
If the model covariances $\cnu$ and $\ceta$ are globally known, and the selection matrix $\St$ is constant for all $t$, a distributed implementation of \eqref{kkfeq} can be derived along the lines of \cite{tvt11}. 
{To this end, notice that substituting~\eqref{iterx} in~\eqref{eq:kkf_estimate}, one can re-write the KKF predictor as
\begin{align}\label{eq:kkf_estimate2}
\yh(t) &= \Sb\left[\b{F}(t) - \b{F}(t)\b{S}(t) \b{K}(t) + \b{K}(t) \right] \yst + \Sb \chih(t-1) \nonumber \\ 
& ~~~~~ + \Sb\left[\b{F}(t) \b{S}(t)\b{K}(t) - \b{K}(t) - \b{F}(t) \right] \b{S}(t)\chih(t-1) 
\end{align}
where $\b{F}(t) := \cnu\Stt\left(\b{S}(t)\cnu\b{S}^T(t) + \sigma^2\b{I}_S\right)^{-1}$. With $\chih(t-1)$ available from the previous iteration, it is clear from~\eqref{eq:kkf_estimate2} that if $\mathbf{d}(t):= \left[\b{F}(t) - \b{F}(t)\b{S}(t) \b{K}(t) + \b{K}(t) \right] \yst$ were available at each node of the network, the KKF predictor~\eqref{eq:kkf_estimate} could be performed locally at each node. Assume that measurements are collected at a sub-set of nodes $\mathcal{V}_s \subset \mathcal{V}$, and node $v \in \mathcal{V}_s$ measures delays of the set of paths $\mathcal{S}_v \subset \mathcal{S}$; that is, $v$ is the end-node of all the paths in $\mathcal{S}_v$. Then, to compute $\mathbf{d}(t)$ in a distributed manner, consider rewriting it as a sum of $|\mathcal{V}_s|$ terms, each involving only the local measurements $\mathbf{y}_{s,v}(t) := [\{y_p| p \in \mathcal{S}_v\}]^T$. Next, collect in the $P \times |\mathcal{S}_v|$ matrix $\mathbf{\Phi}_v(t)$, the columns of matrix $\b{F}(t) - \b{F}(t)\b{S}(t) \b{K}(t) + \b{K}(t)$ corresponding to the paths in $\mathcal{S}_v$. Then, $\mathbf{d}(t)$ can be expressed as $\mathbf{d}(t) = \sum_{v \in \mathcal{V}_s} \mathbf{\Phi}_v(t)\mathbf{y}_{s,v}(t)$, which is equivalent to~\cite{SRG08} 
\begin{subequations}
\label{consensus}
\begin{align}
\{\mathbf{d}_v(t)\}_{v \in \mathcal{V}_s} &= \arg \min_{\{\mathbf{d}_v\}}  \sum_{v \in \mathcal{V}_s} \left\| \mathbf{d}_v - |\mathcal{V}_s|  \mathbf{\Phi}_v(t)\mathbf{y}_{s,v}(t) \right\|_2^2 \\
& \mathrm{s.t. }  ~~ \mathbf{d}_v = \mathbf{d}_{v^\prime} \, , \quad v^\prime \in \bar{\mathcal{V}}_s, v \in \mathcal{V}_s
\end{align}
\end{subequations}
where $\mathbf{d}_v(t)$ represents a local copy of $\mathbf{d}(t)$ at node $v$, and $\bar{\mathcal{V}}_s \subset \mathcal{V}_s$ is the set of nodes communicating with $v$. Building on~\eqref{consensus}, an iterative consensus algorithm whereby each node $v$ exchanges its local copy $\mathbf{d}_v(t)$ only with nodes in $\bar{\mathcal{V}}_s$, can be derived by   
employing the so called alternating direction method of multipliers as detailed in~\cite{SRG08} and \cite{tvt11}. Notice that, since the model covariances are globally known, recursions~\eqref{errormat} can be performed locally at each node.
}
\end{rem}

\subsection{Estimating model parameters} \label{estparam}
The LMMSE-optimal dynamic kriging framework described in Sec. \ref{kriging} requires knowledge of model covariance matrices $\cnu$, $\sigma^2\b{I}_S$, and $\ceta$, to operate. 
Of these, $\sigma^2$ depends on the precision offered by the measurement software, and can be safely assumed known a priori. 

The structure of $\cnu$ is motivated by the modeling assumptions and utilizes topological information. 
Intuitively, propagation, transmission, and processing delays over paths $p, q \in \P$ should be highly correlated if these paths share many links. 
This relationship can be modeled by utilizing the Gramian matrix $\b{G} := \b{R}\b{R}^T$, where $\b{R}$ is the $P \times \abs{\c{E}}$ path-link routing matrix; that is, the $(p,l)$th element of $\b{R}$ is 1 if path $p\in\P$ traverses link $l \in \c{E}$, and 0 otherwise. 
Each off-diagonal entry $(p,q)$ of $\b{G}$ represents the number of links common to the paths $p, q \in \P$.
On the other hand, the elements on the main diagonal of $\b{G}$ count the number of constituent links per path.
The covariance matrix of $\nut$ can therefore be modeled as $\cnu=\gamma\b{G}$. 

A similar model for $\cnu$ was adopted by \cite{nk}, where it was motivated from the property that path delays are sum of link delays, that is, $\nut = \b{R}\xt$, where vector $\xt$ collects the link delays.
Under this assumption, it holds that $\cnu=\gamma\b{G}$ if the link delays are uncorrelated across links, and have covariance matrix $\gamma\b{I}_{\abs{\c{E}}}$. 
{ Note that the KKF and path-selection techniques also work with a generic link-delay covariance matrix $\bs{\Sigma}$, i.e., $\mathbf{C}_\nu = \mathbf{R} \mathbf{\Sigma}\mathbf{R}^T$. 
Unfortunately however, in most IP networks the link delays cannot be directly observed, which makes estimation of $\mathbf{\Sigma}$ difficult, if not impossible.
For example, consider a network (1--2--3) where two end terminals (nodes 1 and 3) are connected via an intermediate router (node 2). Clearly, the delays incurred by the individual links (1--2 and 2--3) cannot be discerned from each other, no matter how accurately the end-to-end delays (between 1 and 3) are measured. 
The same reasoning applies to the corresponding covariance matrices, irrespective of the estimation technique used. }

For the remaining parameters, namely $\gamma$ and $\ceta$, an empirical approach is described next. 
It entails a training phase, and a set of measurements $\{\yst\}_{t = 1}^{t_L}$ collected at time slots $t = 1,\ldots, t_L$. 
During the KKF operation, $t_L - 1$ time slots can be periodically devoted to updating model covariances, while predicting the networks-wide delays $\ybt$ for $t = 1,\ldots, t_L$. 
Let $\widehat{\b{C}}_{\bnu}(t):=\hat{\gamma}(t)\b{G}$ and $\widehat{\b{C}}_{\bs{\eta}}(t)$ denote the estimates of $\cnu$ and $\ceta$, respectively, at time $t$.
Estimating the covariance matrix of the state noise is well-known to be a challenging task, primarily because $\bs{\chi}(t)$ and $\bs{\chi}(t-1)$ are not directly observable. 
Furthermore, methods such as those in~\cite{Mehra70} are not applicable in the present context, as they require the KF to be time-invariant and stationary. 
As shown in~\cite{Myers76}, a viable means of estimating $\ceta$ from $\{\yst\}_{t = 1}^{t_L}$ relies on approximating the noise $\bs{\eta}(t)$ as $\bs{q}(t) := \chih(t) - \chih(t-1)$. 
Then, upon  noticing that the resultant process $\{\bs{q}(\tau)\}$ is temporally-white, the sample mean and covariance of $\bs{q}$ can be obtained as \begin{align}
\hat{\bs{m}}_{\bs{q}}(t_L) &= \frac{1}{t_L-1} \sum_{t = 2}^{t_L} \bs{q}(t) \\
\widehat{\b{C}}_{\bs{q}}(t_L) &= \frac{1}{t_L- 2} \sum_{t = 2}^{t_L}(\bs{q}(t) - \hat{\bs{m}}_{\bs{q}}(t))(\bs{q}(t)-\hat{\bs{m}}_{\bs{q}}(t))^T.
\label{eq:tilde_eta}  
\end{align}
Using~\eqref{eq:tilde_eta}, and exploiting the equality $\Exp\{\widehat{\b{C}}_{\bs{q}}\} = (t_L-1)^{-1} \sum_{t} ( \b{M}(t-1)   - \b{M}(t)) + \ceta$, it follows that an unbiased estimate of $\b{C}_{\b{\eta}}$ can be obtained as 
\begin{align}
\widehat{\b{C}}_{\b{\eta}}(t_L)  & = \widehat{\b{C}}_{\bs{q}}(t_L) + \frac{1}{t_L - 1} \sum_{t = 2}^{t_L} \Big( \b{M}(t) -  \b{M}(t -1) \Big)  \, .
\end{align}

Finally, in order to obtain $\hat{\gamma}$, consider the innovations at time $t$ as $\iota_p(t) := \ypt - \hat{\chi}_p(t-1)$, and notice that if the model covariances are correct, then $\iota_p(t)$ is temporally white and zero-mean \cite{Mehra70}.
Indeed, it is possible to show that $\E{\iota_p(t)\iota_q(t)} = \left[\b{M}(t-1)+\ceta+\cnu\right]_{pq}+\sigma^2$ for any $p,q \in \cSt$~\cite{Myers76}.
Further, let $\c{T}_{pq}:=\{t | 1 \leq t \leq t_L, p,q \in \cSt\}$ be the set of time slots for which paths $p$ and $q$ are both measured. 
Then, the sample covariance between $\iota_p(t)$ and $\iota_q(t)$ is given by $\hat{C}_{pq} := \abs{\c{T}_{pq}}^{-1}\sum_{t \in \c{T}_{pq}}\iota_p(t)\iota_q(t)$ for all pairs $p,q \in \c{P}$.
Given $\b{M}(t-1)$ and $\sigma^2$, this observation yields the following estimate
\begin{align}
\left[\widehat{\b{C}}_{\bnu}(t)\right]_{pq} &= \frac{1}{\abs{\c{T}_{pq}}}\sum_{t\in \c{T}_{pq}}\iota_p(t)\iota_q(t) - \sigma^2 - [\b{M}(t-1) + \widehat{\b{C}}_{\bs{\eta}}(t)]_{pq} \label{eq:sigmahat}.
\end{align}
Indeed, entries of $\widehat{\b{C}}_{\bnu}(t)$ can be updated recursively using $\widehat{\b{C}}_{\bnu}(t-1)$ in \eqref{eq:sigmahat}. 
At each time, only a few entries are updated, depending on which paths are observed (cf. $\cSt$). 

Finally, $\hat{\gamma}(t)$ can be obtained by fitting $\widehat{\b{C}}_{\bnu}(t)$ to $\gamma\b{G}$ in the least-squares sense, which yields
\begin{align}
\hat{\gamma}(t_L) = \frac{\sum_{p,q \in \c{P}}[\b{G}]_{pq}[\widehat{\b{C}}_{\bnu}(t_L)]_{pq}}{\|\b{G}\|_F^2}.
\end{align}

{ As further justification for the random-walk model, it is remarked that a model of the form $\bs{\chi}(t) = \bs{B}\bs{\chi}(t-1) + \bs{\eta}(t)$ requires learning the entries of $\bs{B}$. Since the state vector is not directly observed, estimation of $\bs{B}$ is usually significantly more difficult \cite{AnM79, gg96, gg96-2}. Such a model would also need a longer training phase, and may exhibit poor \emph{generalization performance} if the amount of training data is limited \cite{bishop}. This problem also arises when trying to use the model $\mathbf{C}_\nu = \mathbf{R} \mathbf{\Sigma}\mathbf{R}^T$, where additionally, $\bs{\Sigma}$ is not uniquely identifiable, as explained earlier.}

\section{Online Experimental Design} \label{expdesign}
This section considers the problem of optimally choosing the set of paths $\cSt$ (equivalently, the matrix $\St$) so as to minimize the prediction error. 
To begin with, a simple case is considered where the set $\cSt$ is allowed to contain any $S$ paths. 
Operational requirements may however impose further constraints on $\cSt$, and these are discussed later. 

The prediction error can be characterized by using a scalar function of $\My$; see e.g., \cite{bach}.
To this end, the so called D-optimal design is considered, where the goal is to minimize the function $f(\S(t)):= \log\det(\My)$.
The paths selected at time $t$ are therefore given by the solution of the following optimization problem
\begin{align}
\S^{*}(t) &= \arg\min_{\S\in\P}  f(\S)\label{oed} \\
&\text{s. t.}\hspace{1cm} \abs{\S}= S. \label{budget}
\end{align}
Clearly, tackling \eqref{oed} incurs combinatorial complexity and is challenging to solve exactly, even for moderate-size networks.
Indeed, \eqref{oed} is an example of the so called subset selection problem, which is NP-complete in general; see e.g., \cite{kempe} and references therein.

Interestingly, it is possible to solve \eqref{oed} approximately by utilizing the notion of \emph{submodularity}. 
Consider a function $g(\S)$, which takes as input sets $\S \subset \P$. 
Given a set $\A \in \P$ and an element $p \in \P\setminus\A$, the increment function is defined as $\delta^g_{\A}(p):=g(\A\cup\{p\})-g(\A)$.
Function $g(\cdot)$ is submodular if its increments are monotonically decreasing, meaning $\delta^g_{\A}(p) \geq \delta^g_{\B}(p)$ for all $\A \subset \B \in \P$.
Likewise, $g(\cdot)$ is supermodular iff $\delta^g_{\A}(p) \leq \delta^g_{\B}(p)$ for all $\A \subset \B \in \P$.
In the present case, the following proposition holds.
\begin{prop}\label{propsub}
The function $f(\S)$ is monotonic and supermodular in $\S$.
\end{prop}
The proof of Proposition \ref{propsub} is provided in Appendix \ref{Asub}, and relies on related results from \cite{bach}. 

An important implication of Proposition \ref{propsub} is that a greedy forward selection algorithm can be developed to solve \eqref{oed} approximately \cite{nemh}.
Upon defining the shifted function $h(\S):=f(\S)-\log\det(\b{M}(t-1)+\ceta+\cnu+\sigma^2\b{I}_P)$, a result from \cite{nemh} ensures that the solution of the greedy algorithm $\S^g(t)$ satisfies the inequality
\begin{align}
h(\S^g(t)) \leq \left(1-\frac{1}{e}\right)h(\S^{*}(t)). \label{ratio}
\end{align}
While performance of the greedy algorithm is usually much better in practice, this bound ensures that it does not break down for pathological inputs.

The greedy algorithm involves repeatedly performing the updates $\S \gets \S \cup \arg\min_{p\notin\S}\delta^f_{\S}(p)$ until $\abs{\S}=S$. This is useful in the present case, since the increments can be evaluated efficiently using determinant update rules. Specifically, the updates are given by
\begin{align}
\delta^f_{\emptyset}(p) &= -\log\left(1+\mat{\b{M}(t-1)+\ceta+\cnu}_{p,p}\right) & \forall p \in \P \\
\delta^f_{\S}(p) &= -\log\left(1+\mat{\left((\b{M}(t-1)+\ceta+\cnu)^{-1}+\b{S}^T\b{S}\right)^{-1}}_{p,p}\right) &\forall p\in\P\setminus\S.\label{deltas}
\end{align}
Further, each iteration requires a rank-one update to the matrix inverse in \eqref{deltas}, which can also be performed efficiently. 
The full greedy approach is summarized in Algorithm \ref{greedy}, where $\bs{\Phi}:=(\b{M}(t-1)+\ceta+\cnu)/{\sigma^2}$. 
Algorithm \ref{greedy} involves only basic operations, and it is easy to verify that its worst case complexity is $O(PS^3)$.
Further, the final value of the matrix $\b{V}$ evaluated in the last iteration (Algorithm \ref{greedy}, line 11) is exactly the inverse term required for evaluating the Kalman gain in \eqref{gain}.
In fact, the operational complexity can be further reduced using lazy updates \cite{minoux}. 
{ Finally, it is worth mentioning that the low-complexity of Algorithm 1 is also a result of the random-walk model used here. In particular, the state space model $\bs{\chi}(t) = \bs{B}\bs{\chi}(t-1) + \bs{\eta}(t)$ would result in significantly more complicated expressions. }

\begin{algorithm}
\caption{Greedy algorithm for solving \eqref{oed}}\label{greedy}
\begin{algorithmic}[1]
\vspace{0.2cm}
\algrenewcommand{\algorithmiccomment}[1]{\hskip5em // #1}
\Function{Greedy}{$\bs{\Phi}$, $S$} 
\State $s \gets \arg\max\limits_{1 \leq p\leq P}~[\bs{\Phi}]_{p,p}$
\State $\b{V} := \mat{1/\left([\bs{\Phi}]_{s,s}+1\right)}$
\State $\S \gets \{s\}$
\For{$k=2:S$}
\State $\b{w}_p \gets \bs{\Phi}_{\S,p}$ for all $p \in \P\setminus\S$ \Comment{$\b{w}_p$ has entries $[\bs{\Phi}]_{s,p}$ for all $s\in \S$}
\State $s \gets \arg\max\limits_{p\notin\S}~[\bs{\Phi}]_{p,p} - \b{w}_p^T\b{V}\b{w}_p$ 
\State $\S\gets \S \cup \{s\}$
\State $d \gets [\bs{\Phi}]_{s,s}-\b{w}_s^T\b{V}\b{w}_s+1$
\State $\b{u} \gets -\b{V}\b{w}_s$
\State $\b{V} \gets \mat{\b{V}+\b{u}\b{u}^{T}/d & \b{u}/d \\ \b{u}^T/d & 1/d}$
\EndFor \label{euclidendwhile}
\State \textbf{return} $\S$
\EndFunction
\end{algorithmic}
\end{algorithm}

Next, consider a more practical scenario, where the software installed at each end-node can measure delays on all paths originating at that node.
At any time $t$ however, delays are measured from only $N$ end-nodes. 
Let $\c{V}_e$ denote the set of all end-nodes, and $\P_v$, the set of paths which have the node $v \in \c{V}_e$ as their origin (likewise, $\P_{\N}:= \bigcup_{v \in \N} \P_v$ for $\N \subset \c{V}_e$).
For any subset $\N$ (and its complement $\bar{\N} := \c{V} \setminus \N$), define the selection matrix $\b{N}$ ($\bar{\b{N}}$) consisting of canonical vectors $\b{e}^T_p$ as rows, for all $p \in \P_{\N}$ ($p \in \P_{\bar{\N}}$).
Defining the cost function $f_n(\N):=f(\P_{\N})$, the online optimal design problem for this scenario is expressed as
\begin{subequations}
\begin{align}
\N^{*}(t) &= \arg\min_{\N \subset \c{V}_e} f_n(\N) \label{oedn}\\
&\text{s. t.}\hspace{0.5cm} \abs{\N} = N.
\end{align}
\end{subequations}
It follows from the properties of submodular functions that the cost function $f_n(\N)$ is also monotonic and supermodular in $\N$.
In particular, observe that the increments $\delta_{\N}^n(v) = f_n(\N\cup\{v\})-f_n(\N) = f(\P_{\N} \cup \P_v) - f(\P_{\N})$ for $v \notin \N$ satisfy the non-increasing property, i.e., $\delta_{\A}^n(v) \leq \delta_{\B}^n(v)$  for all $\A\subset\B\subset\c{V}_e$ and $v \notin \B$.
A greedy algorithm similar to Algorithm \ref{greedy} can therefore be developed to obtain an approximate solution with the same $(1-1/e)$ guarantee as in \eqref{ratio}. 
Complexity of the greedy algorithm in this case would be however higher, since evaluating $\delta_{\N}(v)$ now requires rank-$\abs{\P_v}$ updates in the determinant and inverses. 
Nevertheless, the algorithm would still be efficient as long as $\abs{\P_v} \ll P$ for all $v \in \c{V}_e$.
In the special case when delay measurements are performed by only one node per time slot ($N=1$), the solution of \eqref{oedn} is simply given by
\begin{align}
\N^{*}(t) &= \arg\min_{v \in \c{V}_e} \log\det\left(\b{I}_{\abs{\P_v}} + \mat{\b{M}(t-1)+\ceta+\cnu}_{vv}\right)
\end{align}
where $\mat{\b{M}(t-1)+\ceta+\cnu}_{vv}$ is the $\abs{\P_v} \times \abs{\P_v}$ submatrix containing the rows and columns of $\b{M}(t-1)+\ceta+\cnu$ corresponding to the paths in $\P_v$. 

In some networks, it may be relatively straightforward to install delay measurement software on every end-node, while allowing each end-node to measure delay on only one path per time slot.
This amounts to replacing the budget-constraint \eqref{budget} in \eqref{oed} with
\begin{align}
\abs{\S \cap \P_{v}} &= 1 &\forall~v \in \c{V}_e. \label{parts}
\end{align}
Interestingly, constraints of this form can also be handled using the greedy approach by simply imposing \eqref{parts} while searching for the best increment at every iteration.
Specifically, the search space of path $p$ [cf. Algorithm  \ref{greedy}, line 7] now becomes $p \in \P \setminus \P_{\N}$, where $\N=\{v : \S \cap \P_v \neq \emptyset \}$. 
More general constraints of the form $\abs{\S \cap \P_{v}} \leq S_v$ can similarly be incorporated.
Constraints of this form are referred to as partition matroid constraints, under which the greedy algorithm provides an approximation ratio of $1/2$ \cite{fisher}.

\section{Empirical Validation}\label{ev}
Performance of the proposed network-wide latency prediction schemes is validated using two different datasets, which include delays measured on: 
\begin{description}
	\item[(a)] Internet2 backbone network\footnote{[Online] \texttt{http://www.internet2.edu/network}}, a lightly loaded network that exhibits low delay variability; and, 
	\item[(b)] New Zealand Active Measurement Project (NZ-AMP)\footnote{[Online] \texttt{http://erg.cs.waikato.ac.nz/amp}}, a network deployed across several universities and ISPs in New Zealand, characterized by comparatively higher variability in delays.
\end{description}
Using the aforementioned datasets, the performance of KKF is also compared against that of competing alternatives in~\cite{nk} and~\cite{coates}. 

Before proceeding, a brief description of the nonlinear estimation technique in \cite{coates} is provided. The approach hinges on a sparse representation of the network-wide delays, and employs $\ell_1$-norm minimization to recover the sparse basis coefficient vector. Specifically, the path delays adhere to the postulated linear model $\b{y}(t) =  \b{H} \bbeta(t)$, where $\|\bbeta(t)\|_0 \ll P$, and the matrix $\b{H} \in \mathbb{R}^{P \times P}$ is constructed using diffusion wavelets~\cite{Maggioni}. The diffusion matrix used for computing the wavelet basis is obtained by applying Sinkhorn balancing~\cite{sinkhorn} to the matrix $\b{W} \in \mathbb{R}^{P \times P}$, whose $(p,q)$-th element is defined as 
\begin{align}
[\b{W}]_{p,q} &= \frac{[\b{G}]_{pq}}{[\b{G}]_{pp}+[\b{G}]_{qq} - [\b{G}]_{pq}} 
\end{align}
where $\b{G}$ is the Gramian defined in Sec. \ref{estparam}. The overall algorithm amounts to solving the following minimization problem 
\begin{subequations}
\label{wavelets}
\begin{align}
\hat \bbeta^{\prime}(t) & = \arg \min_{\bbeta^{\prime}} \| \bbeta^{\prime} \|_1 \\
\text{s. t.}\hspace{0.5cm} & \yst = \b{S}(t) \b{H} \b{L} \bbeta^{\prime} 
\end{align}
\end{subequations}
where $\b{L}$ is a diagonal matrix whose $(n,n)$-th entry is given by $[\b{L}]_{n,n} = 2^k$, with $k \in \mathbb{N}$ denoting the scale corresponding to the diffusion wavelet coefficient $\beta_n$~\cite{coates}. Subsequently, $\ybt$ is predicted as $\yh(t) = \Sb \b{HL} \hat{\bbeta}^{\prime}(t)$. 

Under the premise that delays change slowly with time, the described algorithm can be used to estimate $\b{y}_{\bar{s}}(t)$ over a sequence of $\tau > 1$ contiguous time-steps jointly. In this case, problem~\eqref{wavelets} is solved by replacing $\yst$ with $\by := [\y_{s}^T(t-\tau+1), \y_{s}^T(t-\tau+2), \ldots, \y_{s}^T(t)]^T$, and by computing the $P \tau \times P \tau$ diffusion wavelet matrix based on $\b{W}$ and temporal correlations as shown in~\cite{coates}. Although this is a viable way to capture temporal correlations of delays, observe that it requires solving $\ell_1$-norm minimization problems with $P\tau$ variables every $\tau$ time slots. 
This increase in complexity prohibits the use of a large value of $\tau$, and the simulations here only report performance with $\tau = 5$.
It is also worth mentioning that such a batch solution also does not compare favorably to a real-time implementation, such as that provided by the KKF where delay predictions become available every time new measurements arrive.

\subsection{Internet2 Delay Data}
\begin{figure}[t]
  \centering
  \includegraphics[width=1.5\figwidth]{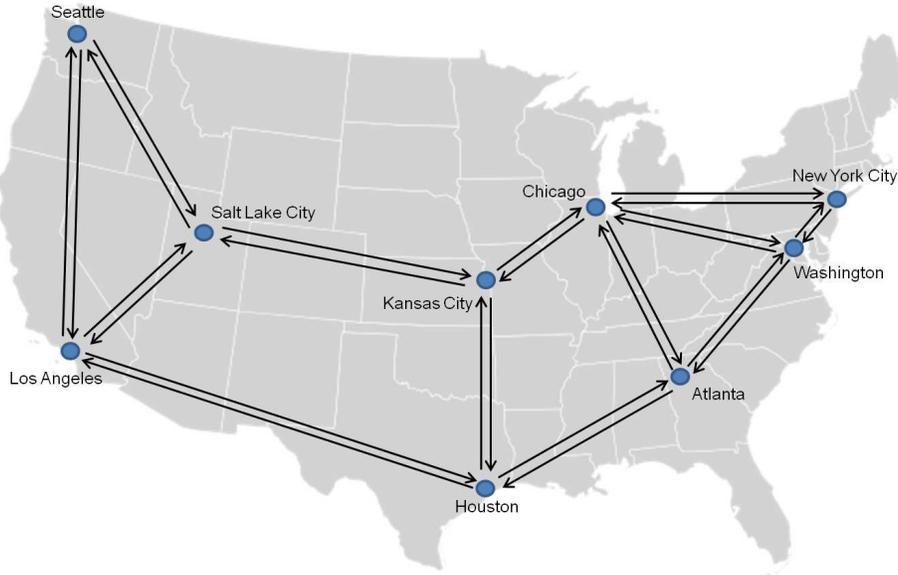}
  \vspace{-.5cm}
  \caption{Internet2 IP backbone network.}
  \label{Internet2}
\end{figure}

The One Way Active Measurement Project (OWAMP) collects one way delays on the Internet2 backbone network\footnote{[Online] \texttt{http://ndb1.net.internet2.edu/cgi-bin/owamp.cgi}}. 
The network has 9 end-nodes and 26 directional links as depicted in Fig.~\ref{Internet2}.  Delays are measured on the 72 paths among the end-nodes every minute. The data $\{\yt\}$ is collected over $t_P = 4500$ minutes (about three days) in July 2011.

The model KKF covariances $\cnu$ and $\ceta$ are estimated using data from the initial 1,000 time slots. In this phase, $50$ paths are randomly selected per time slot. The KKF is initialized by setting $\gamma = 1$, $\ceta = \cnu$, and run for 500 time slots. Next, $\hat{\gamma}(t)$ and $\widehat{\b{C}}_{\b{\eta}}(t)$ are updated in an online fashion, as outlined in Sec. \ref{estparam}. The final values are obtained at the conclusion of the training phase at $t$ = 1,000. 

\begin{figure}
	\centering
  \subfigure[True map\label{dm1}]{\includegraphics[width=0.45\textwidth]{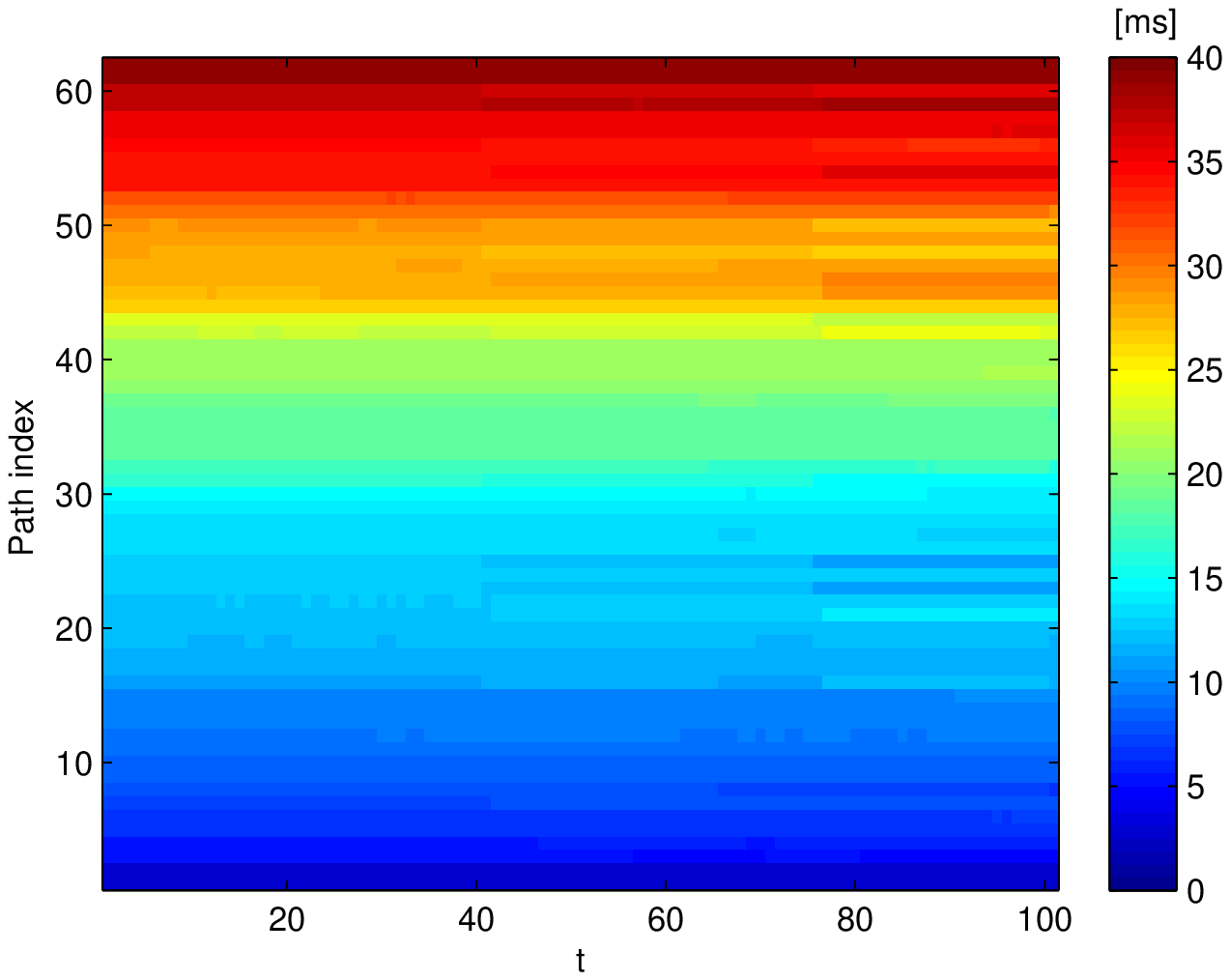}}
  \subfigure[Kriging\label{dm2}]{\includegraphics[width=0.45\textwidth]{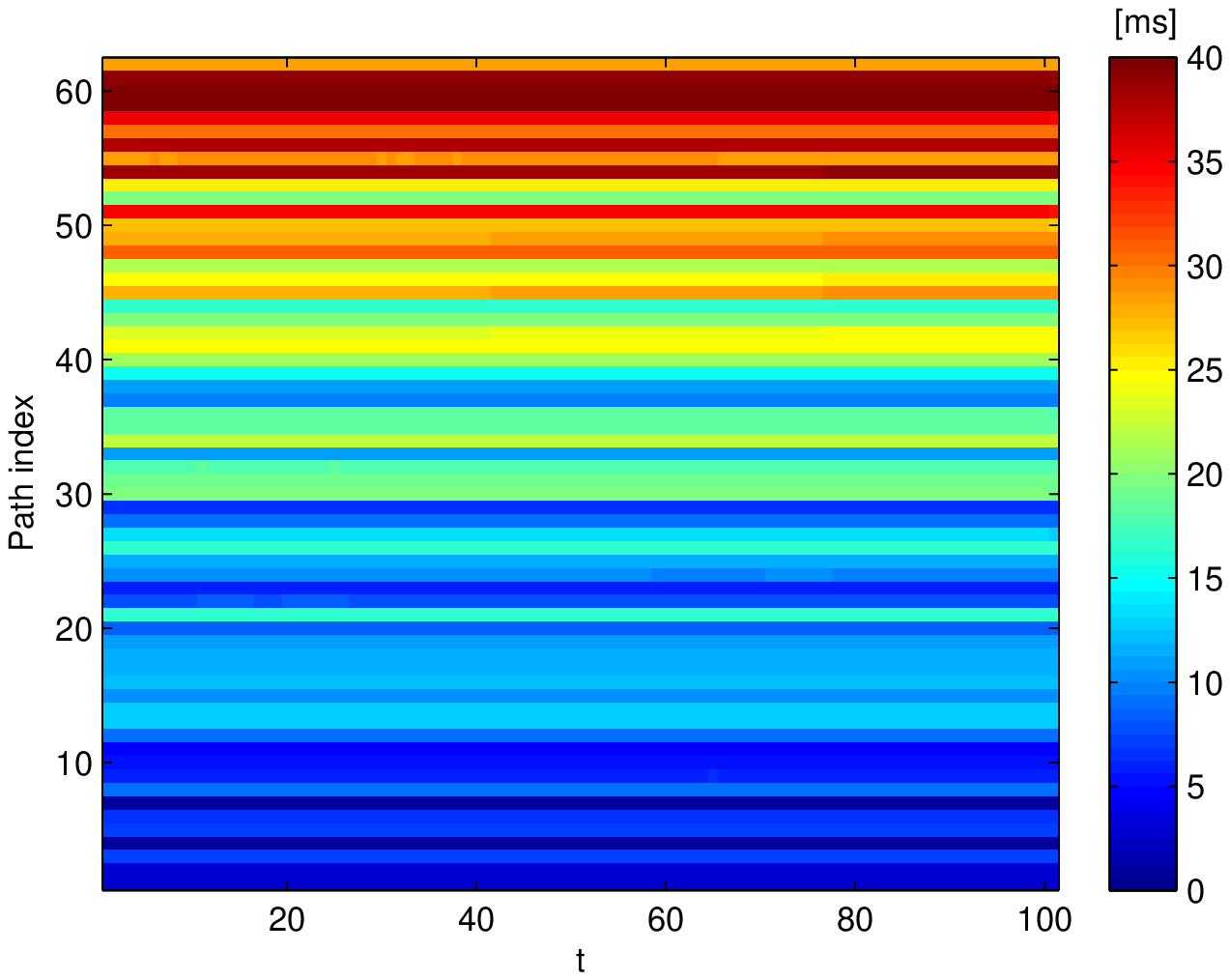}} \subfigure[Wavelets\label{dm3}]{\includegraphics[width=0.45\textwidth]{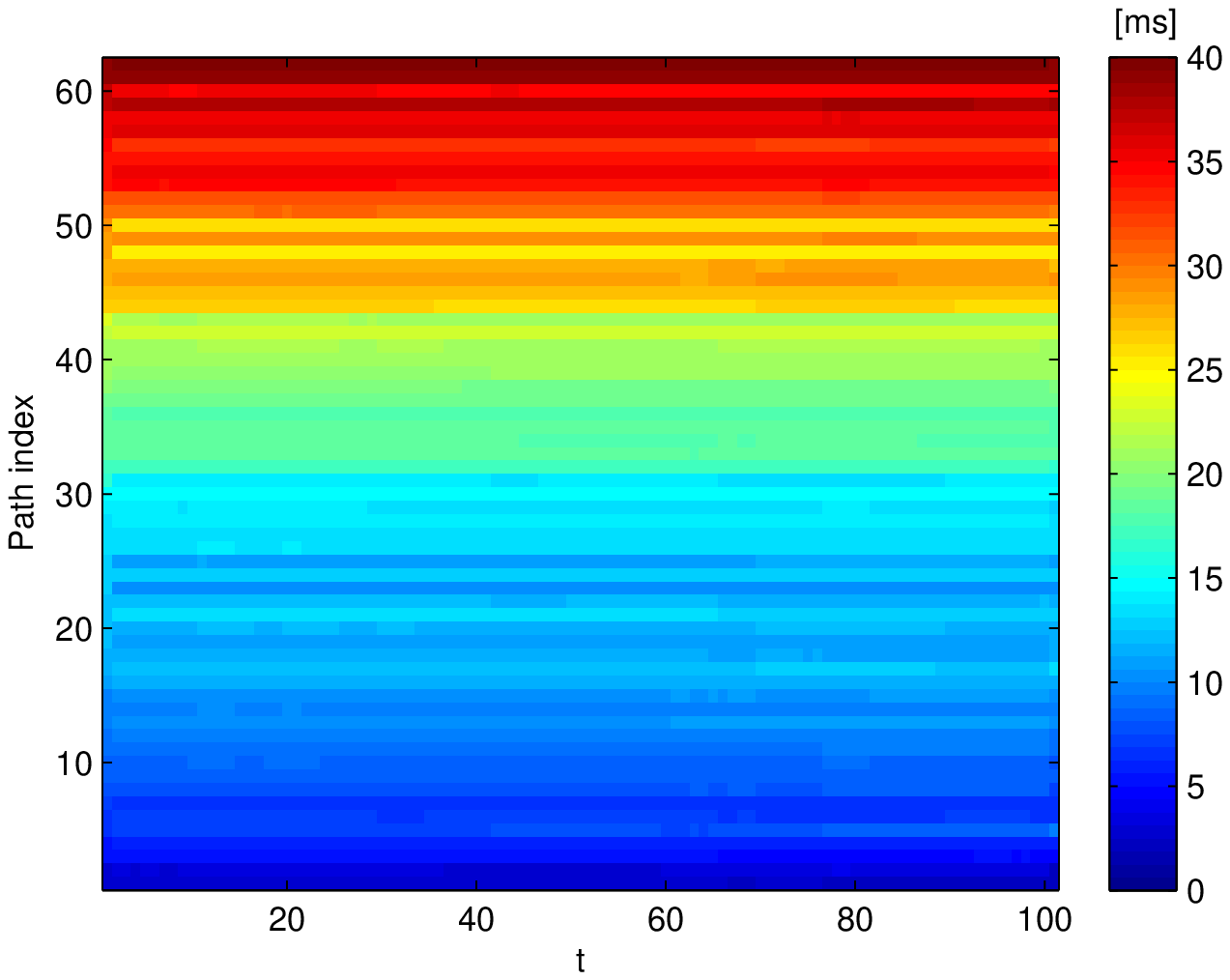}}  \subfigure[KKF\label{dm4}]{\includegraphics[width=0.45\textwidth]{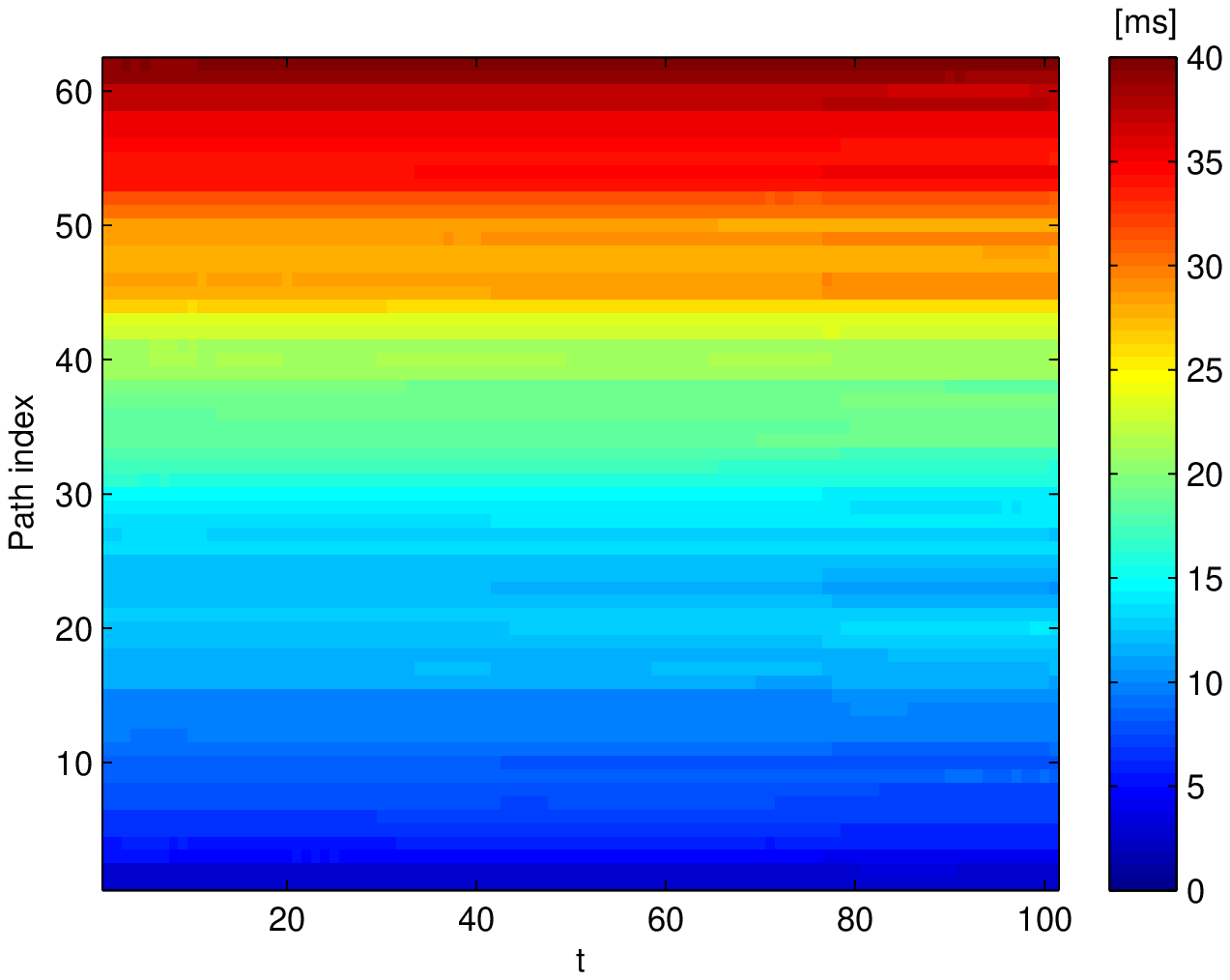}}
  \caption{True and predicted delay map for $62$ paths in the Internet2 network over in interval of $100$ minutes.}
  \label{delaymaps}
\end{figure}

Pictorially, the performance of different algorithms can be assessed through delay maps shown in \ref{delaymaps}. 
Such maps can succinctly represent the network health, and are especially useful for networks which otherwise have low delay variability, such as the Internet2. 
The map in Fig.~\ref{delaymaps}(a) corresponds to the true delays, wheres maps (b), (c), and (d) depict the predicted values obtained from the network kriging, wavelet-based approach, and KKF respectively. 
Predictions are performed using measurements over an interval of $100$ minutes on $10$ random paths (same paths are used throughout the considered interval), and the delays are predicted on the remaining 62 paths are reported. 
In these maps, paths are arranged in increasing order according to the true delay at time $t=1$.
It can be seen that the map produced by the kriging and compressive sensing approaches are very different from the true map.
In contrast, the map obtained when using the KKF is close to the true map. 
In particular, observe that the delays of several paths change slightly around $t=80$ in Fig. \ref{dm1}.
However, of the three maps, this change is only discernible in the KKF map in \ref{dm4}. 
The delay predictions provided by the KKF are thus sufficiently accurate for human inspection at control centers, even when monitoring a few paths. 

\begin{figure}[t]
\centering
\includegraphics[width=1.5\figwidth]{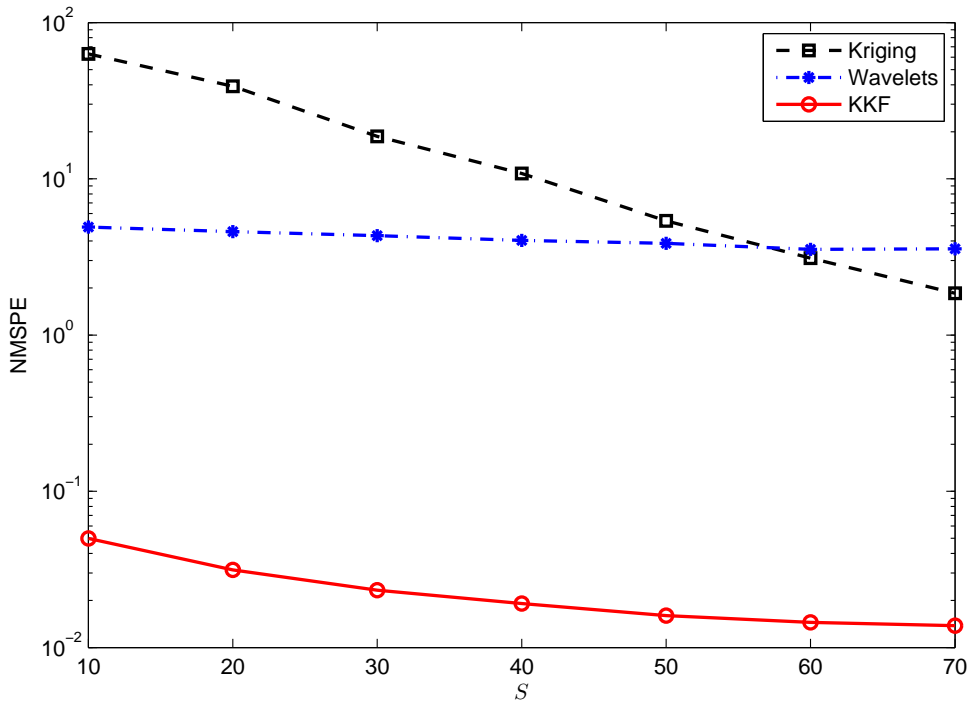}
\caption{NMSPE as a function of $S$, Internet2 network with random path selection.}
\label{mspeI2random}
\end{figure}

{It should be remarked that the maps in Fig. \ref{delaymaps} are only for demonstration purposes, and not much can be inferred about the relative performance of different algorithms from these depictions alone.} For a more detailed analysis of the different delay prediction approaches, consider the normalized mean-square prediction error (NMSPE), defined as
\begin{align}
\text{NMSPE} := \frac{1}{(t_P-t_L)(P-S)}\sum_{t=t_L+1}^{t_P}\left\|\yh(t)-\ybt\right\|_2^2.
\end{align} 
The prediction performance of the three algorithms is first assessed by using delay measurements on randomly selected paths for each $t$. The (same) randomly selected paths are used for all three approaches. 
Fig. \ref{mspeI2random} depicts the NMSPE as a function of $S$, the number of paths on which delays are measured. Clearly, the KKF markedly outperforms the other two approaches across the entire range of $S$. As expected~\cite{coates}, the compressive sampling-based approach provides a more accurate prediction than network kriging. 

\begin{figure}
\centering
\includegraphics[width=1.5\figwidth]{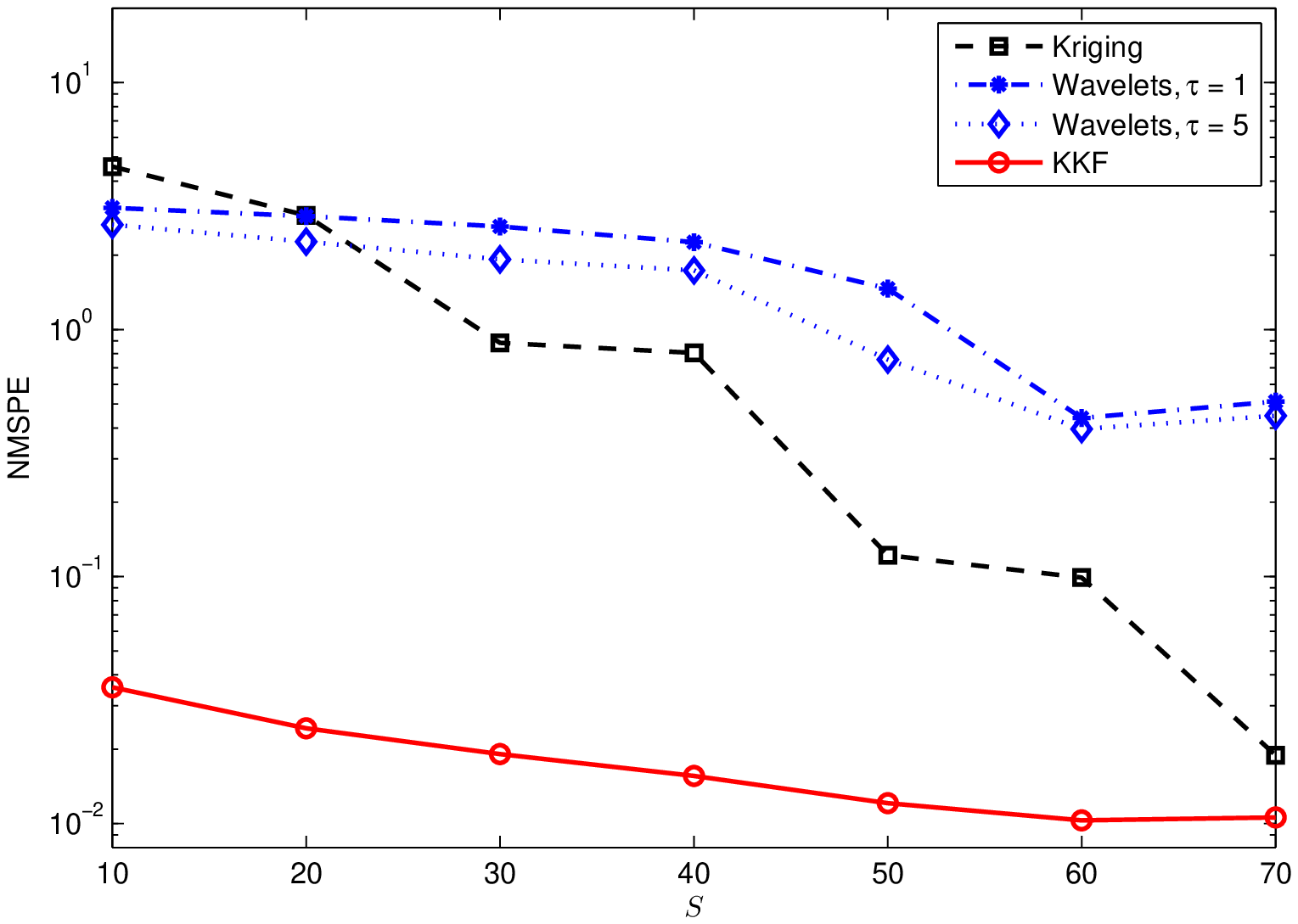}
\caption{NMSPE as a function of $S$, Internet2 network with optimal path selection.}
\label{mspeI2optimal}
\end{figure}

Next, the performance of the three algorithms is analyzed for the case when paths for delay measurement are selected optimally. 
For the network kriging and the wavelet-based approaches, the optimal paths are obtained according to the selection procedures provided in~\cite{nk} and~\cite{coates}, respectively. 
As pointed out in \cite{coates}, performance of the wavelet-based approach can be improved by capitalizing on temporal correlations.
This is done by solving \eqref{wavelets} using measurements from $\tau = 5$ consecutive time slots in a batch form. 
The temporal correlation is set to $0.5$ and the optimal paths are obtained again using the selection strategy outlined in~\cite{coates}. 
For the KKF, optimal paths are selected in an online fashion using Algorithm 1.
Again, a significantly more accurate prediction of the path delays for the entire range of $S$ is obtained via the KKF.

\subsection{NZ-AMP Delay Data}
The KKF algorithm is tested here using delay data from NZ-AMP. 
The project continuously runs \texttt{ICMP} and scamper to determine the topology and delays between a set of nodes in New Zealand. The data collected for this paper consist of end-to-end delays measured every ten minutes over the month of August 2011. The network has a total of 186 paths, whose delays range from almost constant to highly variable, at times reaching up to 250ms.

\begin{figure}
\centering
\includegraphics[width=1.5\figwidth]{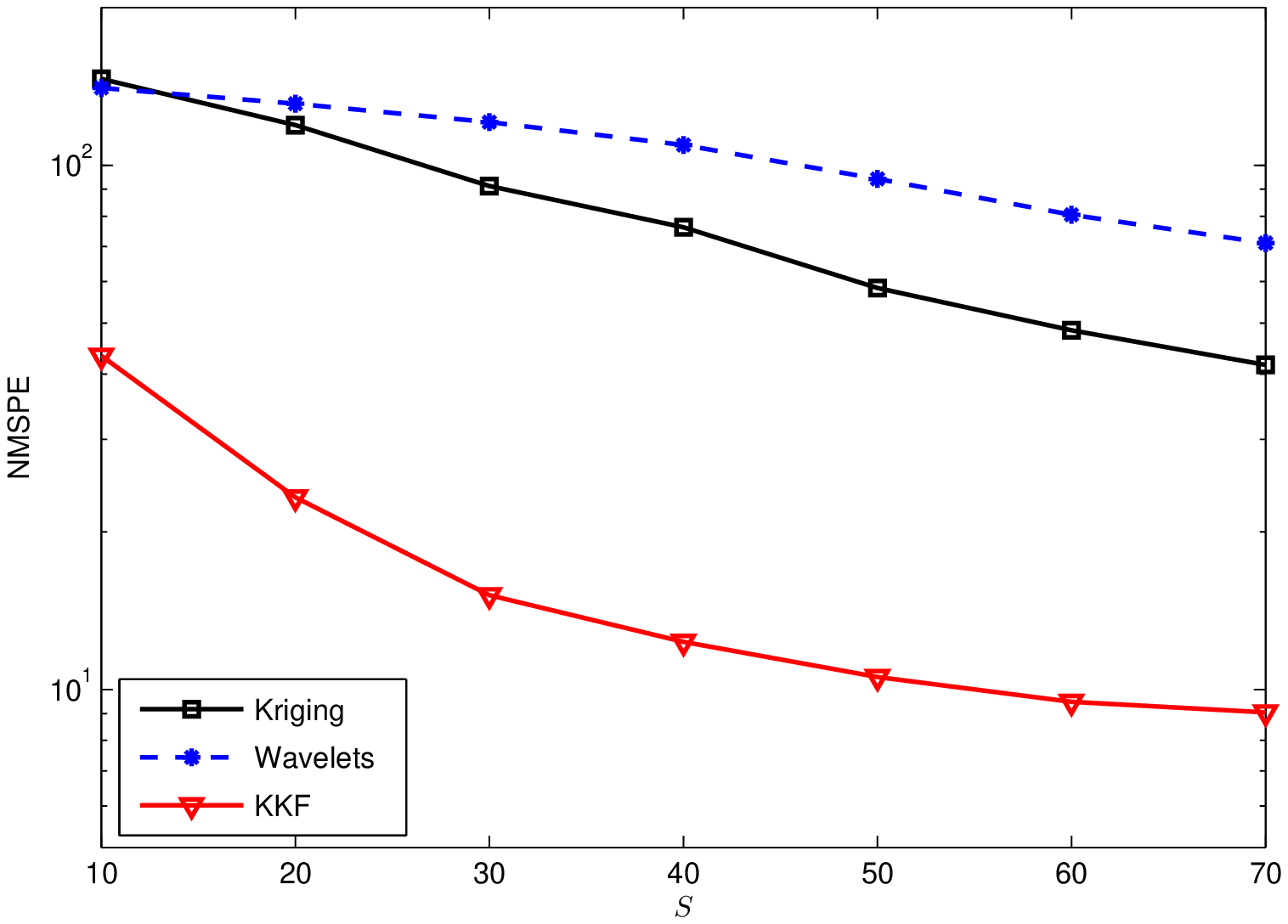}
\caption{NMSPE as a function of $S$, NZ-AMP network with random path selection.}
\label{mspeNZrandom}
\end{figure}

\begin{figure}
\centering
\includegraphics[width=1.5\figwidth]{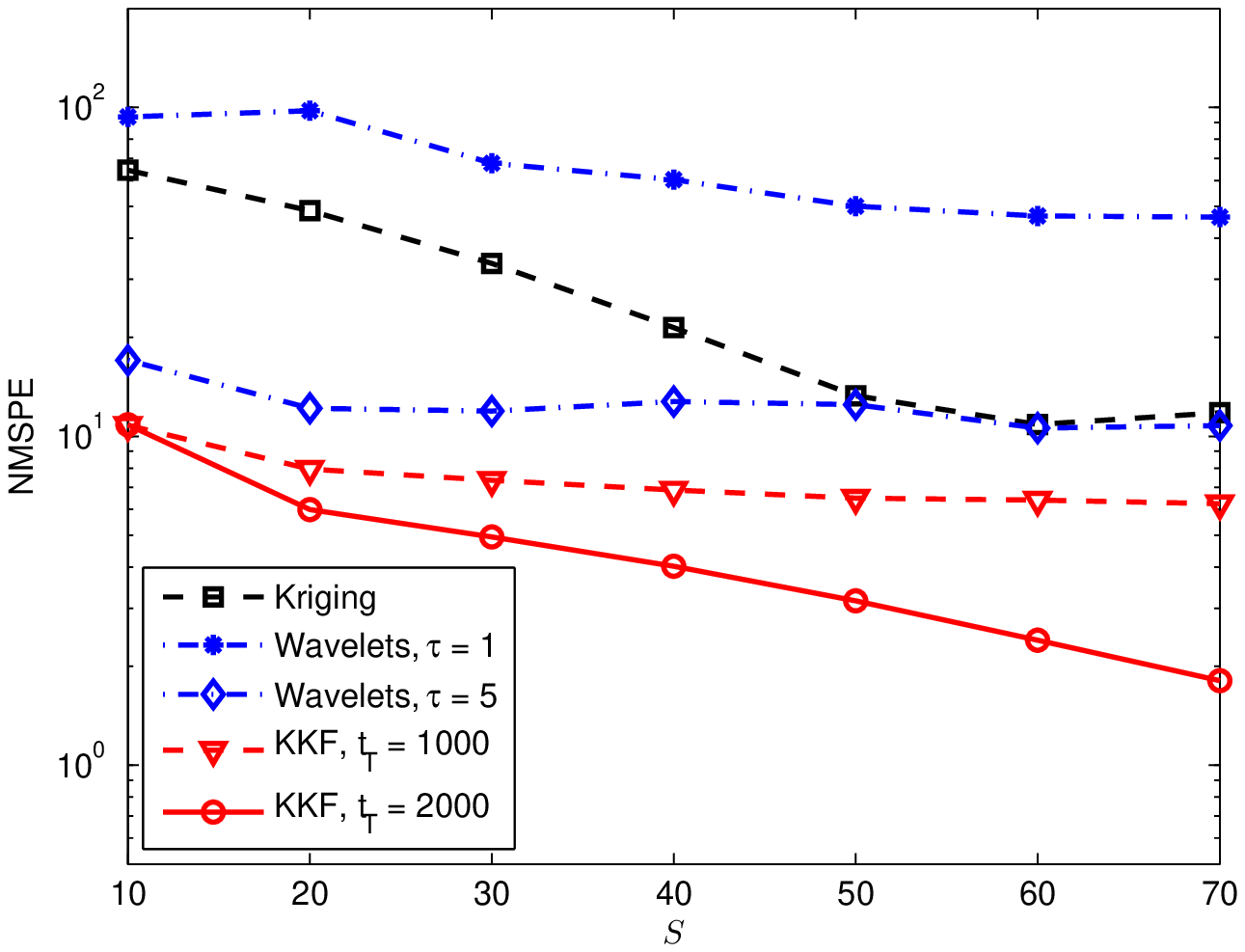}
\caption{NMSPE as a function of $S$, NZ-AMP network with optimal path selection.}
\label{mspeNZoptimal}
\end{figure}

In Fig.~\ref{mspeNZrandom}, the NMSPE as a function of $S$ is reported, for the case where paths that are to be measured are chosen randomly. Again, same paths are used for the three considered  schemes. The KKF provides a markedly lower prediction error also for the NZ-AMP delay data. 
On the other hand, Fig.~\ref{mspeNZoptimal} shows the NMSPE on optimally selected paths for all three schemes. 
The KKF performs relatively better than the competing schemes for this data set as well.
Observe though that the actual values of the NMSPE incurred for this dataset is at least an order of magnitude higher than those in the Internet2 dataset. 
Indeed, given the high variability in the data, it is possible to improve upon the prediction accuracy of KKF by training it better. 
This is showcased by the considerably lower prediction error curve for training interval $t_L$=2,000 shown in Fig.~\ref{mspeNZoptimal}.

\begin{figure}
	\centering
  \subfigure[Kriging\label{sc1}]{
  \begin{tikzpicture}[scale = 0.75]
    \begin{axis} [
        scale only axis,        
        enlargelimits=false,    
        axis on top,            
        xlabel=True delay (ms),
        ylabel=Predicted delay (ms),
    ]
        \addplot graphics [
            xmin=0,
            xmax=300,
            ymin=0,
            ymax=300,
        ] {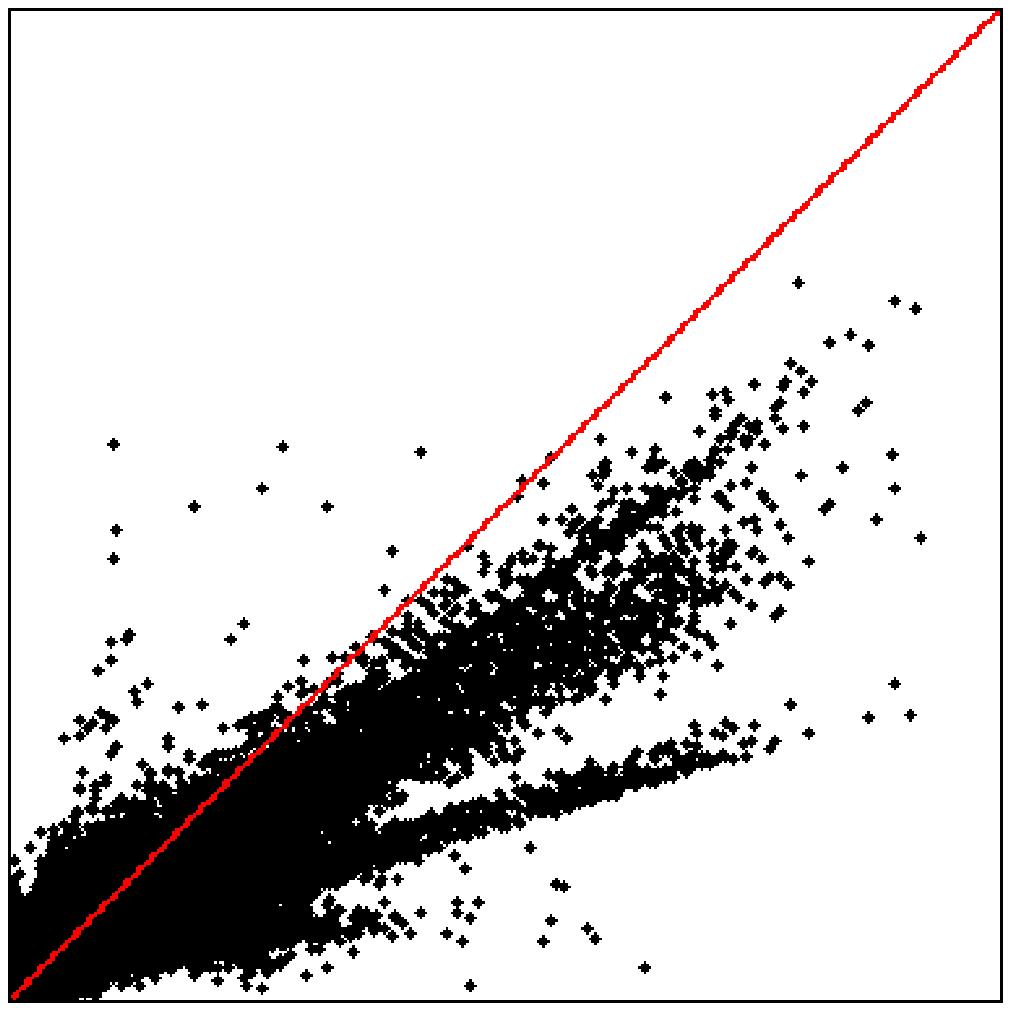};
    \end{axis}
  \end{tikzpicture}}
  \subfigure[Wavelets\label{sc2}]{
  \begin{tikzpicture}[scale = 0.75]
    \begin{axis} [
        scale only axis,        
        enlargelimits=false,    
        axis on top,            
        xlabel=True delay (ms),
        ylabel=Predicted delay (ms),
			]
        \addplot graphics [
            xmin=0,
            xmax=300,
            ymin=0,
            ymax=300,
        ] {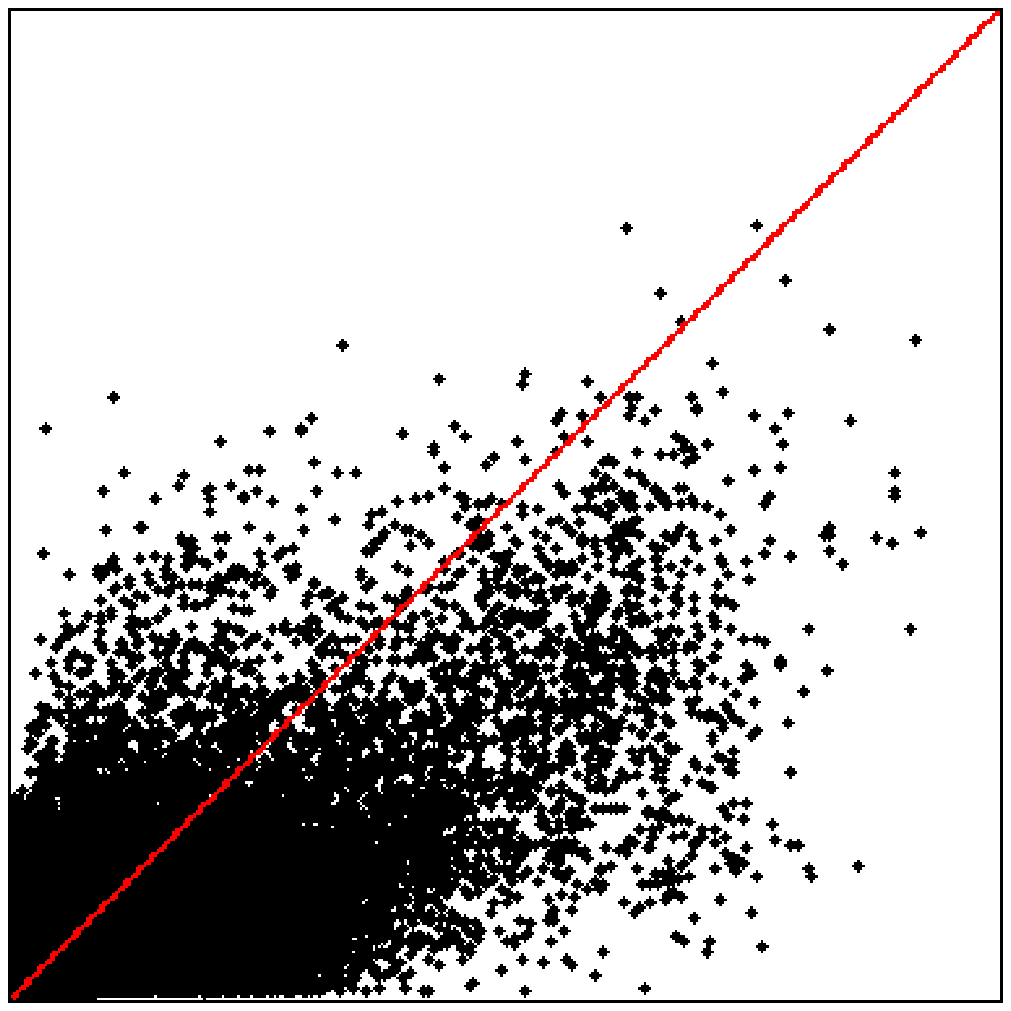};
    \end{axis}
\end{tikzpicture}}  
  \subfigure[KKF \label{sc3}]{
  \begin{tikzpicture}[scale = 0.75]
    \begin{axis} [
        scale only axis,        
        enlargelimits=false,    
        axis on top,            
        xlabel=True delay (ms),
        ylabel=Predicted delay (ms),
            ]
        \addplot graphics [
            xmin=0,
            xmax=300,
            ymin=0,
            ymax=300,
        ] {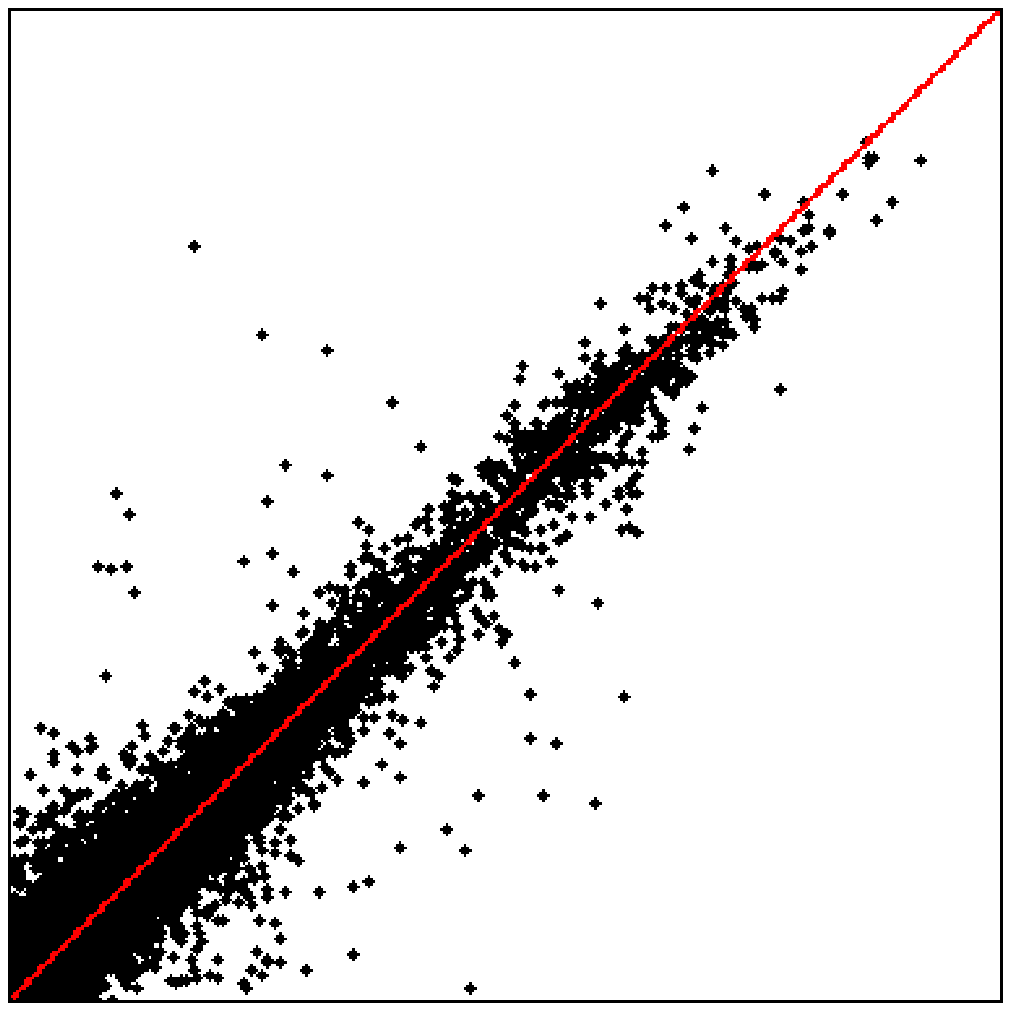};
    \end{axis}
\end{tikzpicture}}
  \caption{Scatter plot for the NZ-AMP network, $S = 30$ with optimal path selection.}
  \label{scatternz}
\end{figure}

While the NMSPE is useful for characterizing the average performance, network operators are also interested in the prediction accuracy over the entire range of delay values.
Towards this end, Fig. \ref{scatternz} shows the scatter plots of $\yh(t)$ versus $\ybt$ for all $t$ and $S = 30$ optimally selected paths.
The points cluster around the $45$-degree line $\yh(t) = \ybt$, and the thinner the ``cloud'' of points is, the more accurate the estimates are. Indeed, it can be seen that the points generated from the KKF estimates are crammed in a very close area around the $45$-degree line, and accurate estimates are produced for the entire range of experienced delays.  Furthermore, the scatter plots corroborate the unbiasedness of the KKF predictor.

\section{Conclusion}
The present paper develops a spatio-temporal prediction approach to track and predict network-wide path delays using measurements on only a few paths.
The proposed algorithm adapts a kriged Kalman filter that exploits both topological as well as historical data.
The framework also allows for the use of submodular optimization in the selection of optimal delay measurement locations.
The problem of path selection is formulated for different types of constraints on the set of selected paths, and solved in an online fashion to near-optimality. 
The resulting predictor is validated on two datasets with different delay profiles, and is shown to substantially outperform competing alternatives.

\appendices

\section{Error covariance matrix}\label{Aerr}
Towards deriving an expression for $\b{M}^{\b{y}}_{\bar{s}}(t)$, observe that the prediction error can be written as 
\begin{align}
\ybt-\yh(t) &= \Sb\chit + \Sb\nut + \epsbt - \Sb\chih(t) \nonumber\\
& \hspace{.5cm} - \Sb\cnu\Stt\left(\St\cnu\Stt + \si\right)^{-1}\left[\yst-\St\chih(t)\right] \\
&= \Sb(\chit - \chih(t) + \nut) + \epsbt \nonumber\\
& \hspace{.5cm} - \Sb\cnu\Stt\left(\St\cnu\Stt + \si\right)^{-1}\left[\St(\chit-\chih(t)+\nut) + \epst\right]. \label{err1}
\end{align}
Using \eqref{iterx}, the term $\chit-\chih(t)$ can be written as
\begin{align}
\chit-\chih(t) &= \chit-\chih(t-1) - \Kt\left[\St(\chit + \nut) + \epst - \St\chih(t-1)\right]\nonumber\\
&= \chit-\chih(t-1) + \Kt\St(\chit-\chih(t-1)+\nut) + \Kt\epst \nonumber\\
&= (\b{I}_P - \Kt\St)\tc - \Kt\St\nut -\Kt\epst \label{err2}
\end{align}
where $\tc := \chit - \chih(t-1)$. Substituting \eqref{err2} in \eqref{err1}, it follows that
\begin{align}
\ybt-\yh(t) &= \Sb(\b{I}_P - \Kt\St)(\tc+\nut) - \Sb\Kt\epst + \epsbt \nonumber\\
&\hspace{0.5cm} - \Sb\cnu\Stt\left(\St\cnu\Stt + \si\right)^{-1} \nonumber\\
&\hspace{1cm} \times \left[\St(\b{I}_P - \Kt\St)(\tc+\nut) - \St\Kt\epst +\epst \right]\\
& \hspace{-0.3cm} = \Sb(\b{I}_P - \Kt\St)(\tc+\nut) \nonumber\\
& \hspace{-0.2cm} - \Sb\cnu\Stt\left(\St\cnu\Stt + \si\right)^{-1} \St(\b{I}_P - \Kt\St)(\tc+\nut)\nonumber\\
& \hspace{-0.2cm} - \Sb\Kt\epst - \Sb\cnu\Stt\left(\St\cnu\Stt + \si\right)^{-1}(\b{I}_S-\St\Kt)\epst \nonumber\\
& + \epsbt 
\end{align}
which, after some manipulations, can be expressed as
\begin{align}
\ybt-\yh(t) &= \Sb(\b{I}_P - \b{Q}(t)\St)(\tc+\nut) + \b{Q}(t)\epst + \epsbt \label{err3}
\end{align} 
where 
\begin{align}
\b{Q}(t):=\Kt + \cnu\St(\St\cnu\Stt+\si)^{-1} - \cnu\St(\St\cnu\Stt+\si)^{-1}\St\Kt.
\end{align}
Next, substituting for $\Kt$ from \eqref{gain}, the expression for $\b{Q}(t)$  simplifies to
\begin{align}
\b{Q}(t) &= (\b{M}(t-1)+\ceta)\Stt\left[\St(\b{M}(t-1)+\ceta+\cnu)\Stt +\si\right]^{-1} \nonumber\\
&\hspace{0.5cm}+ \cnu\Stt(\St\cnu\Stt+\si)^{-1} \nonumber\\ &\hspace{0.5cm}-\cnu\Stt(\St\cnu\Stt+\si)^{-1}\St(\b{M}(t-1)+\ceta)\Stt \nonumber\\
&\hspace{1cm} \times \left[\St(\b{M}(t-1)+\ceta+\cnu)\Stt+\si\right]^{-1} \\
&= (\b{M}(t-1)+\ceta + \cnu)\Stt\left[\St(\b{M}(t-1)+\ceta+\cnu)\Stt+\si\right]^{-1}. \label{qt}
\end{align}
Utilizing the fact that $\tc$, $\nut$, $\epst$, and $\epsbt$ are mutually uncorrelated, with $\E{\tc\tilde{\bs{\chi}}^T(t)}:=\b{M}(t-1)+\ceta$, the error covariance matrix $\My$ becomes
\begin{align}
\My &= \Exp\left[(\ybt-\yh(t))(\ybt-\yh(t))^T\right] \\
&= \Sb(\b{I}_P - \b{Q}(t)\St)(\b{M}(t-1) + \cnu + \ceta)(\b{I}_P - \Stt\b{Q}^T(t))\Sbt \nonumber\\
&\hspace{0.5cm}+ \sigma^2\Sb\b{Q}(t)\b{Q}^T(t)\Sbt + \sib\\
&= \Sb(\b{M}(t-1) + \cnu + \ceta)\Sbt - 2\Sb\b{Q}(t)\St(\b{M}(t-1) + \cnu + \ceta)\Sbt \nonumber\\ 
&\hspace{0.0cm}+ \Sb\b{Q}(t)\St(\b{M}(t-1) + \ceta + \cnu)\Stt\b{Q}^T(t)\Sbt + \sigma^2\Sb\b{Q}(t)\b{Q}^T(t)\Sbt \nonumber\\
&\hspace{1cm}+ \sib\\
&= \Sb(\b{M}(t-1) + \cnu + \ceta)\Sbt - \Sb\b{Q}(t)\St(\b{M}(t-1) + \cnu + \ceta)\Sbt \nonumber\\
&\hspace{1cm}+ \sib. \label{errorfinal}
\end{align}
Substituting for $\b{Q}(t)$ [cf. \eqref{qt}] in \eqref{errorfinal}, and using the Woodbury matrix identity \cite{golub}, the final expression for $\b{M}^{\b{y}}_{\bar{s}}(t)$ becomes
\begin{align}
\My  = \sib + \Sb\left[\big(\b{M}(t-1) + \cnu + \ceta\big)^{-1} + \frac{1}{\sigma^2}\Stt\St\right]^{-1}\Sbt\,.
\end{align}

\section{Proof of monotonicity and supermodularity of $f$}\label{Asub}
Let $\bs{\Phi} := \frac{1}{\sigma^2}(\b{M}(t-1)+\ceta+\cnu)$, and observe that $f$ can be written as
\begin{subequations}
\begin{align}
f(\S) &= \log(\sigma^2) + \log\det\left[\b{I}_{P-S} + \bar{\b{S}}(\bs{\Phi}^{-1}+\b{S}^T\b{S})^{-1}\bar{\b{S}}^T\right] \\
&= \log(\sigma^2) + \log \det \left[\b{I}_P + \bar{\b{S}}^T\bar{\b{S}}(\bs{\Phi}^{-1}+\b{S}^T\b{S})^{-1}\right] \label{syl1}\\
&= \log(\sigma^2) + \log \det \left[\bs{\Phi}^{-1}+\b{S}^T\b{S} + \bar{\b{S}}^T\bar{\b{S}}\right] + \log\det\left[(\bs{\Phi}^{-1}+\b{S}^T\b{S})^{-1}\right]\label{fs}
\end{align}
\end{subequations}
where \eqref{syl1} follows from Sylvester's theorem for determinants \cite{golub}. 

Observing that $\bar{\b{S}}^T\bar{\b{S}} + \b{S}^T\b{S} = \b{I}_P$, it is possible to write $f(\S)$ as
\begin{align}
f(\S) = \log(\sigma^2) + \log\det(\bs{\Phi}^{-1}+\b{I}_P) - \log\det\left(\bs{\Phi}^{-1} + \b{S}^T\b{S}\right).
\end{align}
Next, consider the decomposition $\bs{\Phi} = \b{U}\b{U}^T$, and define the shifted function
\begin{subequations}
\begin{align}
h(\S)&:= f(\S) - \log(\sigma^2) - \log\det\left(\bs{\Phi}+\b{I}_P\right) \\
&= -\log\det(\b{I}_P + \b{S}^T\b{S}\bs{\Phi}) \\
&= -\log\det\left[ \b{I}_S + (\b{S}\b{U})(\b{S}\b{U})^T \right] \label{syl2}
\end{align}
\end{subequations}
where Sylvester's theorem has again been used in \eqref{syl2}. 
Finally, it is well known that a function of the form $\log\det(\b{I}_P + (\b{S}\b{U})^T(\b{S}\b{U}))$ is non-decreasing and submodular (see e.g., \cite{bach}), which allows one to deduce that $f(\S)$ is non-increasing and supermodular. Note further that the greedy approach from \cite{nemh} can be used on $h(\S)$ by defining $h(\emptyset)=0$.

\newpage

\bibliographystyle{IEEEtran}
\bibliography{IEEEabrv,netcart}

\end{document}